# Temperature dependence of normalized sensitivity of Love wave sensor with unidirectional carbon fiber epoxy composite/Mn-doped 0.24PIN-0.46PMN-0.30PT ternary single crystal configuration

Ziqing Luo,[1] Yujiao Ma,[1] Xiaopeng Wang,[1] Naixing Huang,[1,a)] Xudong Qi,[2] Enwei Sun,[3,b)] Rui Zhang,[3] Bin Yang,[3] Tianquan Lü,[3] Jian Liu,[3] and Wenwu Cao[3,4,c)]

[1]*Department of Physics, Northeast Petroleum University, Daqing, Heilongjiang, 163318, China*

[2]*Condensed Matter Science and Technology Institute and Department of Physics, Harbin Institute of Technology, Harbin 150080, China*

[3]*The School of Instrumentation Science and Engineering, Harbin Institute of Technology, Harbin 150080, China*

[4]*Department of Mathematics and Materials Research Institute, The Pennsylvania State University, University Park, PA 16802, USA*

We have derived a general formula for sensitivity optimization of gravimetric sensors and use it to design a high precision and high sensitivity gravimetric sensor using unidirectional carbon fiber epoxy composite (CFEC) guiding layer on single crystal Mn-doped $y$Pb(In$_{1/2}$Nb$_{1/2}$)O$_3$-(1-$x$-$y$)Pb(Mg$_{1/3}$Nb$_{2/3}$)O$_3$-$x$PbTiO$_3$ (Mn:PIN-PMN-PT) piezoelectric substrate. The normalized maximum sensitivity $\left(\left|S_m^f\right|\lambda\right)_{max}$ exhibits a decreasing tendency with temperature up to 55℃. For the CFEC-on-Mn:PIN-PMN-PT sensor configuration with $\lambda$ = 24 μm at 25℃, the maximum sensitivity $\left|S_m^f\right|_{max}$ can reach as high as 760.88 cm$^2$/g, which is nearly twice that of traditional SiO$_2$/ST quartz configuration gravimetric sensor.

Corresponding authors. *E-mail address*: [a)]huangnaixing@163.com (N. Huang), [b)]sunew@hit.edu.cn (E. Sun), [c)]dzk@psu.edu (W. Cao).

Surface acoustic wave sensors have been employed in liquids and gases for immunoassay formats, gas detections, etc.[1-3] It is known that Love waves exist only when a thin layer is attached on a substrate, and the velocity of the shear wave in the layer must be slower than that in the substrate.[4] By analyzing the gravimetric sensors made of fused quartz, polymethylmethacrylate (PMMA), ZnO, $SiO_2$, etc., as the waveguide layers, a general guideline for the improvement of the mass sensitivity has been summarized: the waveguide layer materials should have low shear velocity, low density, and low acoustic attenuation.[5-6] The Love wave device with polymer guiding layers provides not only excellent mass loading sensitivity, but also good temperature stability.[7] In addition, the elastic property of carbon fiber epoxy composites (CFEC) is almost invariable from 25℃ to 55℃.[8] And the CFEC provides acoustic impedance comparable to that of aluminum, which means low acoustic loss.[9] Hence, the unidirectional carbon fiber epoxy composite (CFEC) is selected as the waveguide material in this work. In a Love mode sensor, the insertion loss can be reduced by selecting a substrate with a larger electromechanical coupling coefficient.[10] In resent years, Mn-doped rhombohedral phase ternary single crystal $y$Pb($In_{1/2}Nb_{1/2}$)$O_3$-(1-$x$-$y$)Pb($Mg_{1/3}Nb_{2/3}$)$O_3$-$x$PbTiO$_3$ (Mn:PIN-PMN-PT) has attracted considerable attention for its outstanding piezoelectric properties and high mechanical quality factor.[11] It was report that the ternary PIN-PMN-PT single crystals have higher phase transition temperatures ($T_{RT}$ and $T_C$), which are advantageous for higher temperature applications.[12-15] The Mn:PIN-PMN-PT ternary single crystal has a lower pyroelectric coefficient, lower relative permittivity, and smaller dielectric loss compared to the binary ones.[16] In addition, the Mn:PIN-PMN-PT single crystals possess high piezoelectric constants and electromechanical coefficients,[17] which help enhance the efficiency of a transducer in converting electrical to mechanical energy (and vice versa). For these reasons, this work uses Mn-doped 0.24PIN-0.46PMN-0.30PT ternary single crystal as the substrate material of the gravimetric sensor. In order to design high precision and high sensitivity gravimetric sensors with temperature compensation, it is important to know the relationship between the normalized sensitivity and environmental temperature.

Partial wave theory is the most commonly used method for analyzing acoustic wave propagation in anisotropic media.[18-21] In this work, we derived the dispersion equation by employing partial wave theory for a specific sensor structure using Mn-doped PIN-PMN-PT ternary single crystal poled along $[001]_c$ pseudo-cubic direction as the substrate. Based on modal analysis, optimal parameters have been derived for designing a gravimetric sensor with a CFEC-on-Mn:PIN-PMN-PT configuration. This letter reports the characterization of the temperature dependence for the normalized sensitivity from 25℃ to 55℃ and optimal design for the normalized waveguide thickness of the sensor.

The wave guiding layer and piezoelectric substrate are bonded as shown in Fig.1. It is a basic structure supporting Love waves without loading layer. The guiding layer is unidirectional CFEC with the fibers are parallel to the $x_2$-axis, which is perpendicular to the wave propagation direction. The substrate is Mn-doped 0.24PIN-0.46PMN-0.30PT ternary single crystal poled along $[001]_c$ pseudo-cubic direction. It is assumed that the $[100]_c$ pseudo-cubic direction of the single crystal is along the $x_1$ direction. The elastic wave equations in the media are given by

$$\rho \frac{\partial^2 u_j}{\partial t^2} - c_{ijkl} \frac{\partial^2 u_k}{\partial x_i \partial x_l} - e_{kij} \frac{\partial^2 \varphi}{\partial x_i \partial x_k} = 0, \tag{1a}$$

$$e_{ikl} \frac{\partial^2 u_k}{\partial x_i \partial x_l} - \varepsilon_{ik} \frac{\partial^2 \varphi}{\partial x_i \partial x_k} = 0 \quad (i,j,k,l=1,2,3), \tag{1b}$$

where $\rho$ is the density, $c_{ijkl}$ is the elastic stiffness tensor component at constant electric field, $e_{kij}$ is the piezoelectric coefficient tensor component at constant strain, and $\varepsilon_{ik}$ is the dielectric permittivity tensor component at constant strain. The particle displacements and the electric potential in the media can be written as

$$u_j = \alpha_j \exp(ikbx_3)\exp[ik(x_1 - vt)], \tag{2a}$$

$$\varphi = \alpha_4 \exp(ikbx_3)\exp[ik(x_1 - vt)], \tag{2b}$$

where $v$ is the phase velocity, $k$ is the magnitude of wave vector **k**, $b$ is the decay coefficient along $x_3$, $\alpha$'s give the relative amplitudes. Since the guiding layer thickness $h$ is finite and the substrate is considered infinite half-space (the thickness of substrate is much greater than the wavelength), the general solutions of displacements and the electric potential can be written as[22]

$$\hat{u}_j = \left\{\sum_n C_n \alpha_j^{(n)} \exp(ikb^{(n)} x_3)\right\} \exp[ik(x_1 - vt)], \quad (n=1,2,\cdots,8), \tag{3a}$$

$$\hat{\varphi} = \left\{\sum_n C_n \alpha_4^{(n)} \exp(ikb^{(n)} x_3)\right\} \exp[ik(x_1 - vt)], \quad (n=1,2,\cdots,8), \tag{3b}$$

$$u_j = \left\{\sum_m C_m \alpha_j^{(m)} \exp(ikb^{(m)} x_3)\right\} \exp[ik(x_1 - vt)], \quad (m=a,b,c,d), \tag{3c}$$

$$\varphi = \left\{\sum_m C_m \alpha_4^{(m)} \exp(ikb^{(m)} x_3)\right\} \exp[ik(x_1 - vt)], \quad (m=a,b,c,d), \tag{3d}$$

where, $C_n$ and $C_m$ are weighting factors of these partial waves in the guiding layer and substrate, respectively. The symbols marked with '^' are parameters of the waveguide layer to distinguish them from that of piezoelectric substrate.

The particle displacements and the three components of stress must be continuous across the interface ($x_3 = 0$) assuming no slippery interface. The electrical boundary conditions involve the continuity of the electric potential and the normal component of the electric displacement across both the surface and the interface. At the surface ($x_3 = h$) of the basic structure, the traction must vanish. In the coordinate system shown in Fig. 1, the boundary conditions of the problem are as follows:

$$u_1 = \hat{u}_1, u_2 = \hat{u}_2, u_3 = \hat{u}_3, T_{31} = \hat{T}_{31}, T_{32} = \hat{T}_{32}, T_{33} = \hat{T}_{33}, \varphi = \hat{\varphi}, D_3 = \hat{D}_3, (x_3 = 0), \quad (4a)$$

$$\hat{T}_{31} = 0, \hat{T}_{32} = 0, \hat{T}_{33} = 0, \hat{D}_3 = k\varepsilon_0 \hat{\varphi}, (x_3 = h). \quad (4b)$$

The traction stresses are expressed by

$$T_{3j} = c_{3jkl}(\partial u_k / \partial x_l) + e_{k3j}(\partial \varphi / \partial x_k), \quad (5)$$

and the normal component of the electric displacement is given by

$$D_3 = e_{3kl}(\partial u_k / \partial x_l) - \varepsilon_{3k}(\partial \varphi / \partial x_k). \quad (6)$$

When being poled along [001]$_c$, the Mn-doped 0.24PIN-0.46PMN-0.30PT ternary single crystal has a 4mm macroscopic symmetry. The CFEC-on-Mn:PIN-PMN-PT structure fits the NP53 symmetry.[22] The [001]$_c$ poled Mn-doped 0.24PIN-0.46PMN-0.30PT ternary single crystal is used as the substrate, and the wave propagation is along the $x_1$ direction in our design. It should be pointed out that the unidirectional CFEC is not elastically isotropic. Substituting Eq. (2) and material constants from Table I & Ref. [24] into Eq. (1), and using Eqs. (3)–(6), we can derive the dispersion relation of Love waves:

$$\tan\left\{kh\hat{\delta}\left[(v/\hat{v}_s)^2-1\right]^{1/2}\right\}=\eta\frac{\delta}{\hat{\delta}}\left[\frac{1-(v/v_s)^2}{(v/\hat{v}_s)^2-1}\right]^{1/2}, \qquad (7)$$

where $\hat{v}_s < v < v_s$ (for exact real solutions of $k$), $\eta = c_{44}/\hat{c}_{44}$, $\delta = (c_{66}/c_{44})^{1/2}$, $\hat{\delta} = (\hat{c}_{55}/\hat{c}_{44})^{1/2}$, $v_s = \delta(c_{44}/\rho)^{1/2}$, $\hat{v}_s = \hat{\delta}(\hat{c}_{44}/\hat{\rho})^{1/2}$, $k = 2\pi/\lambda$. Using Eqs. (1)–(7), the distributions of Love mode displacements, which are normalized to the values at the surface ($x_3 = h$), can be expressed as:

$$U_{(x_3)}^{(1)}=\cos\left\{k\hat{\delta}\left[(v/\hat{v}_s)^2-1\right]^{1/2}h\right\}\exp\left\{k\delta\left[1-(v/v_s)^2\right]^{1/2}x_3\right\}, \quad (x_3 \leq 0), \qquad (8)$$

$$U_{(x_3)}^{(2)}=\cos\left\{k\hat{\delta}\left[(v/\hat{v}_s)^2-1\right]^{1/2}(x_3-h)\right\}, \quad (0 \leq x_3 \leq h). \qquad (9)$$

In practical device configurations, a thin loading layer (density $\rho_3$, shear modulus $\mu_3$, shear wave velocity $v_{S3}$ and thickness $d$) is generally loaded on the surface of the guiding layer. Based on the perturbation theory,[25] the relative frequency sensitivity definition for a gravimetric sensor is given by[4-5,9]

$$S_m^f = \frac{1}{f_0}\lim_{\Delta m_s \to 0}\left(\frac{\Delta f}{\Delta m_s}\right)=\frac{1}{\rho_3 d}\frac{\Delta f}{f_0}=-\frac{(1-v_{S3}^2/v_0^2)}{2\hat{\rho}\Gamma(h)}, \qquad (10)$$

where $\Delta f = f - f_0$, $f$ and $f_0$ are the oscillation frequencies at perturbed and unperturbed cases, respectively, $\Delta m_s$ is the mass per unit area, and $\Gamma(h)$ is given by

$$\Gamma(h)=\int_h^0\left|U_{(x_3)}^{(2)}\right|^2 dx_3+\frac{\rho}{\hat{\rho}}\int_0^{-\infty}\left|U_{(x_3)}^{(1)}\right|^2 dx_3. \qquad (11)$$

When the Rayleigh hypothesis is applicable, i.e., when the elasticity of the loading layer can be ignored $(v_{S3}^2 = \mu_3/\rho_3 = 0)$, the sensitivity equation becomes:

$$S_m^f = \frac{-1}{2\hat{\rho}\Gamma(h)}. \qquad (12)$$

Substituting Eqs. (8)-(9) into Eq. (11) then into Eq. (12) gives the final sensitivity equation:

$$S_m^f = \frac{-1}{\hat{\rho}h}\left(1 + \frac{\sin(\beta_2 h)\cos(\beta_2 h)}{\beta_2 h} + \frac{\rho}{\hat{\rho}}\frac{\cos^2(\beta_2 h)}{\beta_1 h}\right)^{-1}, \qquad (13)$$

with $\beta_1 = k\delta\left[1-(v/v_s)^2\right]^{1/2}$, $\beta_2 = k\hat{\delta}\left[(v/\hat{v}_s)^2-1\right]^{1/2}$, and $k = 2\pi/\lambda = 2\pi f/v$.

The dispersion curves of Love waves for an unidirectional CFEC guiding layer on Mn:PIN-PMN-PT ternary single crystal piezoelectric substrate basic structure at 25℃ are shown in Fig. 2. Each dispersion curve starts at the shear horizontal (SH) wave velocity (2809.9 m/s) of the Mn:PIN-PMN-PT ternary single crystal. At this speed, all higher order Love modes have low-frequency cutoffs. At high frequencies, the phase velocities of all Love modes approach asymptotically to the shear wave velocity 1916 m/s in CFEC. At other temperatures (*T* = 30, 35, 40, 45, 50, 55℃), we can obtain similar dispersion curves. However, it must be pointed out that the SH wave velocity (2761.1 m/s at 55℃) in the substrate decreases with temperature, while the low-frequency cutoffs of higher order Love modes have the rising trend.

Fig. 3 shows the normalized sensitivity $|S_m^f|\lambda$ for the first four modes of the CFEC/Mn:PIN-PMN-PT configuration at 25℃. We find that the first-order mode possesses the highest $\left(|S_m^f|\lambda\right)_{max}$. For the first-order mode, the optimal ratio of the waveguide thickness to the wavelength is $(h/\lambda)_{opt}$ = 0.2691 the normalized maximum sensitivity is $\left(|S_m^f|\lambda\right)_{max}$ = 1.8261 ($10^{-3}$m$^3$/kg) which is nearly twice that of traditional fused quartz/ST quartz structure sensor.[4] Since the wavelength used in Ref. [26] was 24 μm, for the CFEC/Mn:PIN-PMN-PT configuration at 25℃, the optimal design parameters lead to: $h$ = 6.46 μm and the maximum sensitivity of $\left|S_m^f\right|_{max}$ = 760.88 (cm$^2$/g) which is a much higher sensitivity than 400 cm$^2$/g for a traditional structure.

Fig. 4 shows the normalized sensitivity $\left|S_m^f\right|\lambda$ for the first-order mode at different temperatures. The peak data from Fig. 4 are listed in Table Ⅱ, which reveal that the normalized maximum sensitivity $\left(\left|S_m^f\right|\lambda\right)_{max}$ decreases with temperature, meanwhile, the optimal ratio $(h/\lambda)_{opt}$ exhibits a slow increasing tendency, while both elastic constants $c_{44}^E$ & $c_{66}^E$ exhibit a decreasing tendency.

The data points shown in Fig. 4 are the simulation results of the normalized maximum sensitivity and optimal ratio of normalized layer thickness at different temperatures. The solid lines in Fig. 5 are fitted results, which can be described by the following equations:

$$\left(\left|S_m^f\right|\lambda\right)_{max} = 1.85722 - 7.11431 \times 10^{-4} T - 1.90938 \times 10^{-5} T^2 \ (10^{-3} m^3/kg) \qquad (14)$$

$$(h/\lambda)_{opt} = 0.25978 + 4.18431 \times 10^{-4} T - 2.27476 \times 10^{-6} T^2 \qquad (15)$$

From Figs. 4-5 & Table Ⅱ, one can find that the changing trend of normalized maximum sensitivity $\left(\left|S_m^f\right|\lambda\right)_{max}$ is strongly dependent on the change of elastic constants $c_{44}^E$ & $c_{66}^E$ of the piezoelectric substrate with temperature. Though the optimal design $(h/\lambda)_{opt}$ has a slight increasing trend, it is approximately invariant from 25℃ to 55℃. At the optimal design points $(h/\lambda)_{opt}$, the velocity values ($v_{opt}$) of

the first-order mode Love waves have a decreasing tendency with increasing temperature as shown in Table II. And for a given thickness $h$, the sensitivity can be always raised by increasing the operation frequency in a specified sensor configuration,[4] obviously for a fixed wavelength (e.g. 24 μm), the operation frequency values at the optimal points have a decreasing trend with temperature. Because the unidirectional CFEC can be fabricated by different carbon proportions, some changes of the elastic constants, density, acoustic impedance and thermal stability could be beneficial for improving the performance of the sensor. In addition, the design changes of the fibers direction may bring some benefits. Therefore, the CFEC is a very promising waveguide material and the high electromechanical coefficient of Mn:PIN-PMN-PT ternary single crystals with good temperature stability certainly make it an ideal substrate.

This research was supported in part by the NSFC under Grant No. 11304061 and 51572056.

# References

[1]H. Wohltijon and R. Ressy, Anal. Chem. **51**, 1458 (1979).

[2]M. Puiu, A. M. Gurban, L. Rotariu, S. Brajnicov, C. Viespe, and C. Bala, Sensors **15**, 10511 (2015).

[3]W. Wang, S. Y. Fan, Y. Liang, S. T. He, Y. Pan, C. H. Zhang, and C. Dong, Sensors **18**, 3247 (2018).

[4]B. Jakoby and M. J. Vellekoop, Smart Mater. Struct. **6**, 668 (1997).

[5]Z. Wang, J. D. N. Cheeke, and C. K. Jen, Appl. Phys. Lett. **64**, 2940 (1994).

[6]J. Du and G. L. Harding, Sens. Actuators A. **65**, 152 (1998).

[7]W. Wang, X. Xie, J. L. Hou, and S. T. He, IEEE Sensors, Taipei, Taiwan, pp.1-4, (2012).

[8]S. Sathish, J. Welter, R. Reibel, and C. Buynak, AIP Conf. Proc. **820**, 1015 (2006).

[9]N. X. Huang, T. Q. Lü, R. Zhang, and W. W. Cao, Appl. Phys. Lett. **103**, 053507 (2013).

[10]J. S. Liu and S. T. He, J. Appl. Phys. **107**, 073511 (2010).

[11]W. C. Ou, S. Y. Li, W. W. Cao, and M. Yang, J. Electroceram. **37**, 121 (2016).

[12]Y. M. Zhou, Q. Li, F. P. Zhuo, Q. F. Yan, Y. L. Zhang, and X. C. Chu, Ceram. Int. **44**, 9045 (2018).

[13]Y. Chen, K. H. Lam, D. Zhou, Q. W. Yue, Y. X. Yu, J. C. Wu, W. B. Qiu, L. Sun, C. Zhang, H. S. Luo, H. L. W. Chan, and J. Y. Dai, Sensors **14**, 13730 (2014).

[14]S. J. Zhang and F. Li, J. Appl. Phys. **111**, 031301 (2012).

[15]S. J. Zhang, F. Li, N. P. Sherlock, J. Luo, H. J. Lee, R. Xia, R. J. Meyer, Jr., W. Hackenberger, and T. R. Shrout, J. Cryst. Growth **318**, 846 (2011).

[16]Y. Li, Y. X. Tang, J. W. Chen, X. Y. Zhao, L. R. Yang, F. F. Wang, Z. Zeng, and H. S. Luo, Appl. Phys. Lett. **112**, 172901 (2018).

[17]X. Q. Huo, S. J. Zhang, G. Liu, R. Zhang, J. Luo, R. Sahul, W. W. Cao, and T. R. Shrout, J. Appl. Phys. **113**, 074106 (2013).

[18]C. W. Chen, R. Zhang, H. Chen, and W. W. Cao, Appl. Phys. Lett. **91**, 02907 (2007).

[19]A. H. Nayfeh and H. T. Chien, J. Acoust. Soc. Am. **91**, 1250 (1992).

[20]C. H. Yang and D. E. Chimenti, Appl. Phys. Lett. **63**, 1328 (1993).

[21]C. H. Yang and D. E. Chimenti, J. Acoust. Soc. Am. **97**, 2103 (1995).

[22]G. W. Farnell and E. L. Adler, in *Physical Acoustics*, edited by W. P. Mason and R. N. Thurston (Academic, New York, 1972) Vol. **IX**, p. 35.

[23]N. X. Huang, T. Q. Lü, R. Zhang, and W. W. Cao, Chin. Phys. B., **23** (11), 117704 (2014).

[24]L. G. Tang, H. Tian, Y. Zhang, and W. W. Cao, Appl. Phys. Lett. **108**, 082901 (2016).

[25]B. A. Auld, *Acoustic Fields and Waves in Solids* (Wiley, New York, 1976), Vol. **2**, Chap. 12, p. 277.

[26]F. Moreira, M. E. Hakiki, F. Sarry, L. L. Brizoual, O. Elmazria, and P. Alnot, IEEE Sensors J. **7**, 336 (2007).

TABLE I. Material constants of the unidirectional carbon fiber epoxy composite (CFEC) for the fibers are parallel to the $x_2$-axis.

| Elastic constants ($10^{10}$ N/m$^2$) | | | | | | | Density (kg/m$^3$) |
|---|---|---|---|---|---|---|---|
| $c_{11}$ | $c_{12}$ | $c_{22}$ | $c_{23}$ | $c_{44}$ | $c_{55}$ | $c_{66}$ | $\rho$ |
| 1.37 | 0.67 | 0.71 | 12.6 | 0.58 | 0.33 | 0.58 | 1580 |

[a]Ref. [23]

TABLE Ⅱ. Normalized maximum sensitivity $\left(\left|S_m^f\right|\lambda\right)_{max}$ and optimal design of normalized layer thickness ($h/\lambda$) for the first-order mode device with unidirectional CFEC/Mn:doped 0.24PIN-0.46PMN-0.30PT configuration at different temperatures.

| Temperature (℃) | $\left(\left|S_m^f\right|\lambda\right)_{max}$ ($10^{-3}$ m$^3$/kg) | $(h/\lambda)_{opt}$ | $v_{opt}$ (m/s) | $c_{66}^E$ ($10^{10}$ N/m$^2$)[a] | $c_{44}^E$ ($10^{10}$ N/m$^2$)[a] |
|---|---|---|---|---|---|
| 25 | 1.8261 | 0.2691 | 2492.0 | 6.409 | 6.171 |
| 30 | 1.8204 | 0.2701 | 2488.2 | 6.390 | 6.137 |
| 35 | 1.8099 | 0.2710 | 2484.0 | 6.353 | 6.101 |
| 40 | 1.7981 | 0.2731 | 2476.5 | 6.312 | 6.063 |
| 45 | 1.7858 | 0.2744 | 2471.1 | 6.269 | 6.023 |
| 50 | 1.7722 | 0.2752 | 2467.1 | 6.222 | 5.982 |
| 55 | 1.7618 | 0.2757 | 2464.2 | 6.188 | 5.930 |

[a]Ref. [24]

**Figure captions**

Fig. 1. Illustration of the sensor design and the coordinate system.

Fig. 2. Dispersion curves of Love waves at 25℃.

Fig. 3. Normalized sensitivity $\left|S_m^f\right|\lambda$ for the first four modes at 25℃.

Fig. 4. Normalized sensitivity $\left|S_m^f\right|\lambda$ for the first-order mode device with unidirectional CFEC/Mn-doped 0.24PIN-0.46PMN-0.30PT configuration at different temperatures.

Fig. 5. Fitted curves of normalized maximum sensitivity $\left(\left|S_m^f\right|\lambda\right)_{max}$ and optimal design of normalized layer thickness ($h/\lambda$) for the first-order mode device with unidirectional CFEC/Mn-doped 0.24PIN-0.46PMN-0.30PT configuration from 25℃ to 55℃.

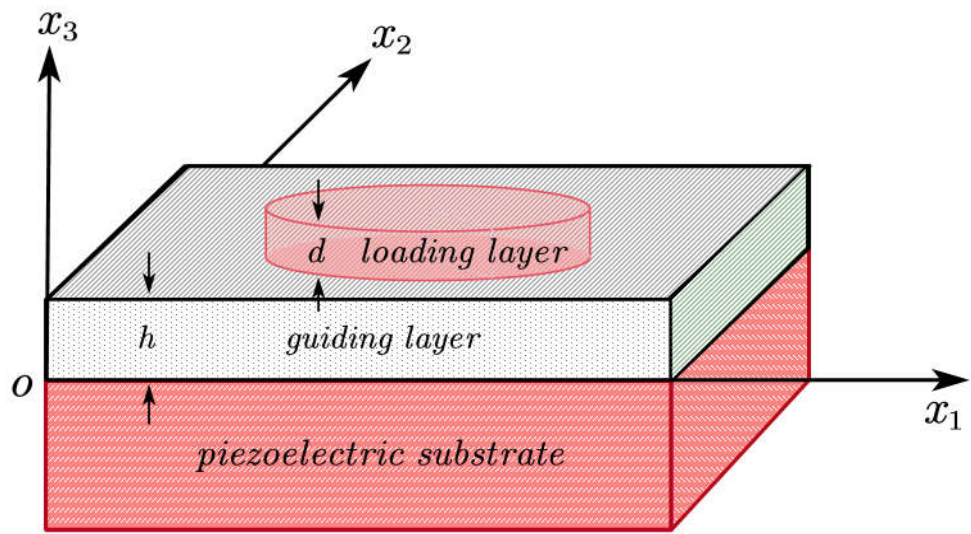

Fig. 1.

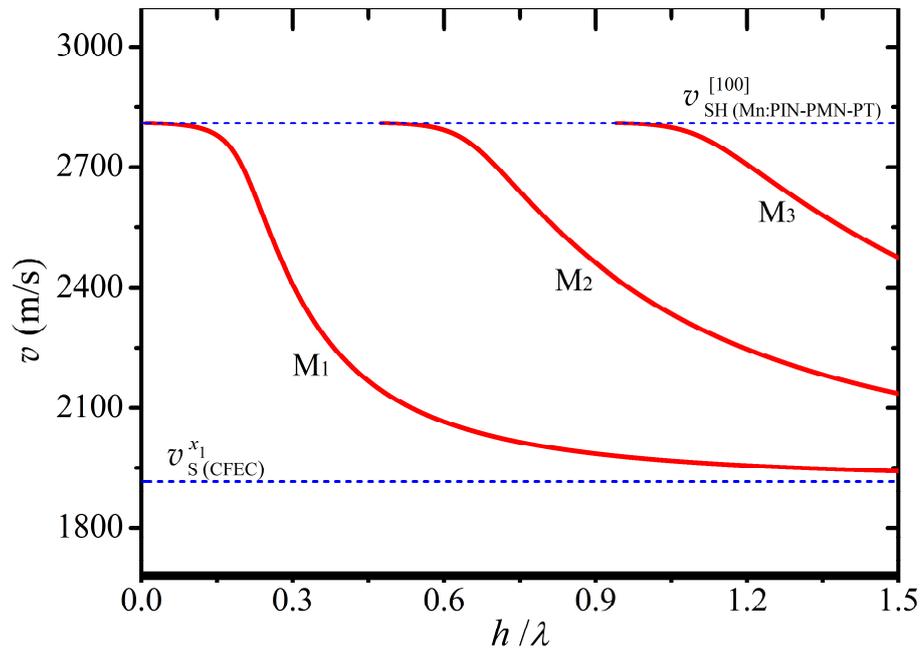

Fig. 2.

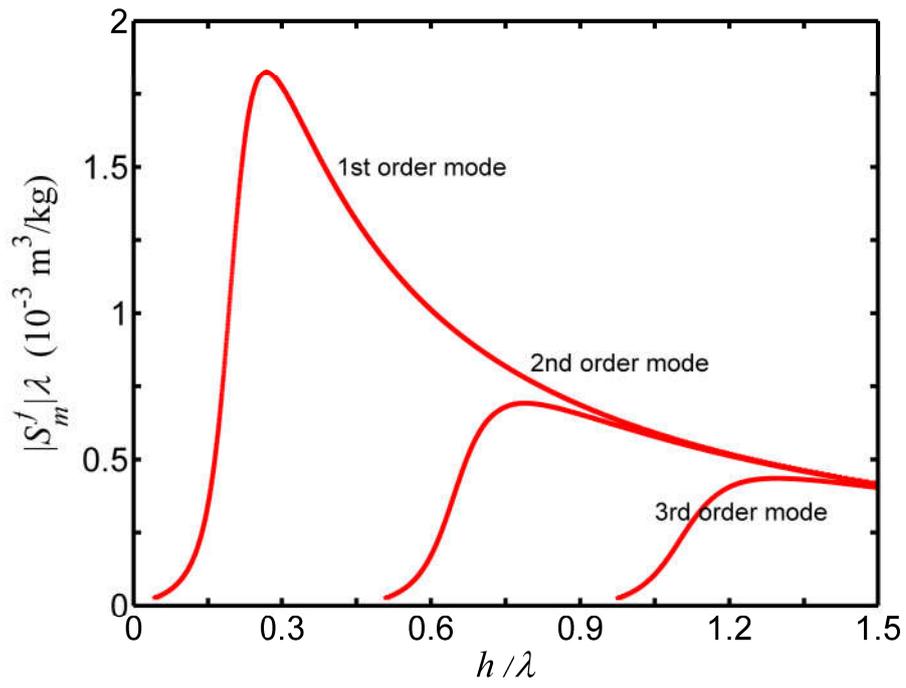

Fig. 3.

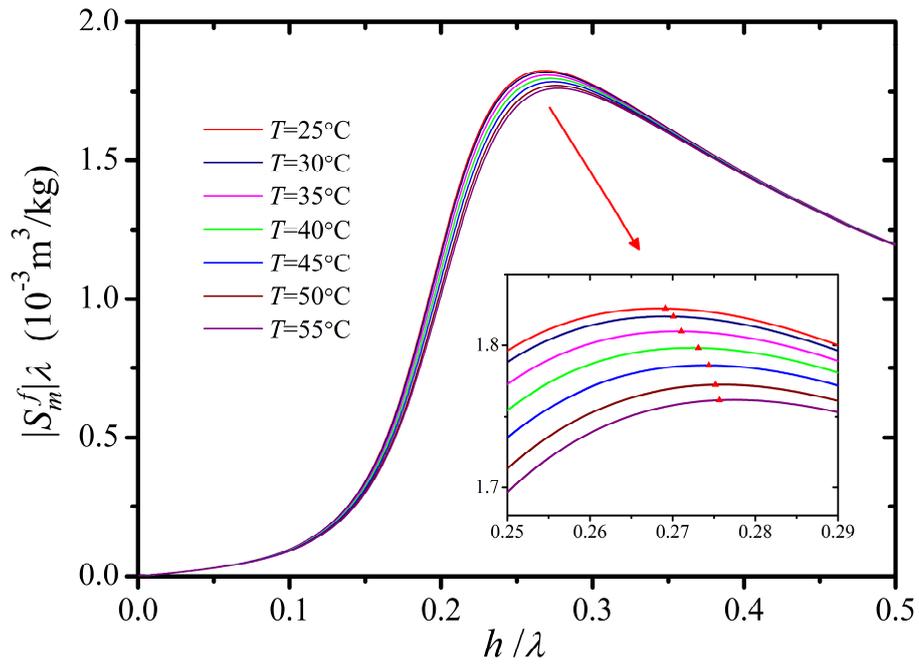

Fig. 4.

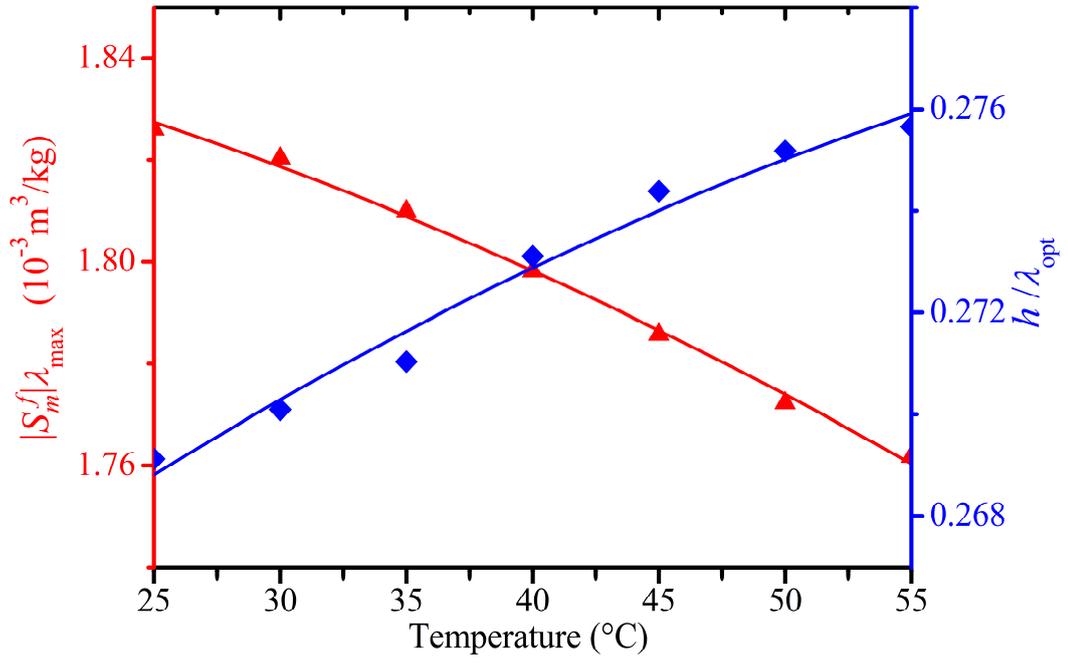

Fig. 5.